\documentclass[twocolumn]{jpsj2} 
\title{Antiferromagnetic fluctuations and the Fulde-Ferrell-Larkin-Ovchinnikov state\\ in CeCoIn$_5$}

\author{M. \textsc{Nicklas}$^{1}$\thanks{E-mail address: nicklas@cpfs.mpg.de}, C. F. \textsc{Miclea}$^{1}$, J. L. \textsc{Sarrao}$^{2}$, J. D. \textsc{Thompson}$^{2}$, G. \textsc{Sparn}$^{1}$ and F. \textsc{Steglich}$^{1}$}

\inst{$^{1}$Max Planck Institute for the Chemical Physics of
Solids, Dresden, Germany \\
$^{2}$Los Alamos National Laboratory, Los Alamos, New Mexico, USA}

\abst{The heavy-fermion superconductor CeCoIn$_5$ is the first
material, where different experimental probes show strong evidence
pointing to the realization of the Fulde-Ferrell-Larkin-Ovchinnikov
(FFLO) state. The inhomogeneous superconducting FFLO state with a
periodically modulated order parameter was predicted to appear in
Pauli-limited, sufficiently clean type-II superconductors already
more than 40 years ago. On the other hand, CeCoIn$_5$ is supposed to
be close to a magnetic quantum critical point (QCP) showing strong
antiferromagnetic (AFM) spin fluctuations (SF) at atmospheric
pressure. We studied the evolution of the FFLO phase away from the
influence of the strong AFM-SF by heat capacity experiments under
pressure ($0\,{\rm GPa} \leq P \leq 1.5\,{\rm GPa}$, $0\,{\rm T}
\leq \mu_0H \leq 14\,{\rm T}$, and $100\,{\rm mK} \leq T \leq
4\,{\rm K}$). Our results prove the stability of the the FFLO phase
under pressure. It even expands, while the Pauli-limiting becomes
weaker and the AFM-SF are suppressed. This shows the intriguing
influence of the AFM-SF on the FFLO state.}

\kword{superconductivity, high pressure, heavy fermion, FFLO state}

\begin{document}
\maketitle

\section{Introduction}

Recently, the possible realization of a
Fulde-Ferrell-Larkin-Ovchinnikov (FFLO) superconducting (SC) state
in CeCoIn$_5$ attracted lot of
attention.\cite{Radovan03,Bianchi03FFLO} The FFLO state is a
spatially inhomogeneous SC phase with a periodically modulated
order parameter.\cite{Fulde64,Larkin65} It was predicted to appear
in clean-limit type-II superconductors close to the upper critical
field, $H_{c2}(0)$, if the orbital pair breaking is small relative
to the Pauli-limiting effect.\cite{Gruenberg66} Furthermore, it
has been shown that a low-dimensional electronic structure and
$d$-wave symmetry of the SC order parameter reinforces the
stability range of the FFLO state.\cite{Shimahara94,Yang98}
CeCoIn$_5$ fulfills all these conditions: (i) CeCoIn$_5$ is in the
clean limit;\cite{Movshovich01} (ii) the Pauli-limiting exceeds
the orbital-limiting effect, indicated by a Maki
parameter\cite{Maki64} $\alpha=\sqrt{2}H_{orb}/H_P>1.8$ ($H_{\rm
orb}$ and $H_{\rm P}$ are the orbital- and the Pauli-limiting
field, respectively);\cite{Bianchi03FFLO} (iii) the electronic
structure is highly anisotropic, dominated by warped cylindrical
Fermi-sheets; \cite{Settai01} (iv) superconductivity in CeCoIn$_5$
is of $d$-wave type, most likely of $d_{x^2-y^2}$ symmetry, based
on results of specific heat\cite{Movshovich01,Aoki04}, thermal
conductivity\cite{Movshovich01,Izawa02}, NMR relaxation
rate,\cite{Kohori01,Yashima04} penetration
depth,\cite{Ormeno02,Chia03} and Andreev-reflection
measurements.\cite{Park05,Rourke05}. This makes CeCoIn$_5$ a
favored candidate for the realization of the FFLO state, and
indeed different physical properties indicate an anomaly inside
the SC state, taken as the transition from the Abrikosov vortex
state to the FFLO
phase.\cite{Radovan03,Bianchi03FFLO,Capan04,Watanabe04,Martin05}

At atmospheric pressure, CeCoIn$_5$ shows the hallmarks of
non-Fermi-liquid (NFL) behavior expected in the vicinity of a
magnetic instability. In the normal state close to the upper
critical field (i) the Sommerfeld-coefficient, $\gamma$, is
diverging logarithmically,\cite{Petrovic01,Kim01,Bianchi03QCP} (ii)
the electrical resistivity $(\rho-\rho_0)\propto
T$,\cite{Petrovic01,Bianchi03QCP,Paglione03,Ronning05} and (iii) the
spin-lattice relaxation rate, $1/T_1\propto
T^{\frac{1}{4}}$\cite{Kohori01,Kawasaki03}. These dependencies point
to the vicinity of an antiferromagnetic (AFM) quantum critical point
(QCP). In detailed studies of electrical resistivity and specific
heat as a function of temperature in different magnetic fields
($H>H_{c2}(0)$) the characteristic temperature,  $T_{FL}$, below
which Landau-Fermi-liquid (LFL) behavior is recovered appears to
vanish at a critical field, $H_{\rm QCP}$, close to $H_{c2}(0)$
suggesting the existence of a magnetic
QCP.\cite{Bianchi03QCP,Paglione03,Ronning05} Recent experiments show
that CeCoIn$_5$ becomes AFM on slight Cd-doping supporting the
proximity to a $H=0$ AFM-QCP.\cite{Pham06} This is corroborated by
the comparison of the pressure-temperature ($P-T$) phase diagrams of
CeRhIn$_5$ and CeCoIn$_5$, which leads to the conclusion that
CeCoIn$_5$ is close to a AFM-QCP at a small, slightly negative
pressure.\cite{Hegger00,Llobet04,Nicklas01,Sidorov02,Knebel04}

The SC FFLO state and the spin fluctuations (SF), possibly AFM,
emerging in the vicinity of the field-tuned magnetic QCP close to
$H_{c2}(0)$ appear in, or even {\it share}, only a small part of the
$H-T$ phase diagram. The aim of this paper is to study the mutual
connection of the AFM-SF and the FFLO state.

\section{Experimental}

In general, in Ce-based inter metallic compounds the application of
hydrostatic pressure suppresses magnetism and eventually a
non-magnetic state is achieved at high pressures. In CeCoIn$_5$
hydrostatic pressure strongly suppresses the AFM-SF. Electrical
resistivity\cite{Sidorov02} and specific heat\cite{Lengyel02}
studies under pressure indicate that the AFM-SF are already
significantly suppressed at $P\gtrsim1.5$~GPa, and LFL is recovered
at low temperatures. Additional evidence for the recovery of the LFL
state comes from pressure dependent de Haas-van
Alphen\cite{Shishido03} and NMR experiments.\cite{Yashima04}
Therefore, heat capacity measurements under pressure, at low
temperatures and in magnetic fields are the especially suited to
study the evolution of the $H-T$ phase diagram with pressure and to
investigate the FFLO phase away from the influence of strong
magnetic fluctuations.

The heat capacity experiments were carried out on high-quality
single crystals of CeCoIn$_5$ grown from excess In-flux. CeCoIn$_5$
crystallizes in the tetragonal HoCoGa$_5$ crystal
structure\cite{Grin79} that can be viewed as layers of CeIn$_3$ and
CoIn$_2$ units stacked sequentially along the
$c$-axis.\cite{Petrovic01} A miniature Cu-Be piston-cylinder-type
pressure cell was utilized to generate pressures up to 1.5 GPa. In a
dilution cryostat, a quasi-adiabatic heat-pulse technique for
temperatures down to 100 mK and magnetic fields up to 12 T was
employed, while a relaxation method was used in a commercial
Physical Property Measurement System (Quantum Design) for magnetic
fields up to 14 T and temperatures $0.350 {\rm ~K} \leq T\leq4 {\rm
~K}$. In all experiments the magnetic field was either applied in
the $ab$-plane or parallel to the $c$-direction of the tetragonal
structure. The narrow width of the SC transition of lead which
served as pressure gauge/medium confirmed the good quasi-hydrostatic
pressure conditions inside the pressure cell. Special care was taken
to ensure a precise orientation of the sample with respect to
magnetic field $H$.

\begin{figure}[b]
\begin{center}
\includegraphics[angle=0,width=85mm,clip]{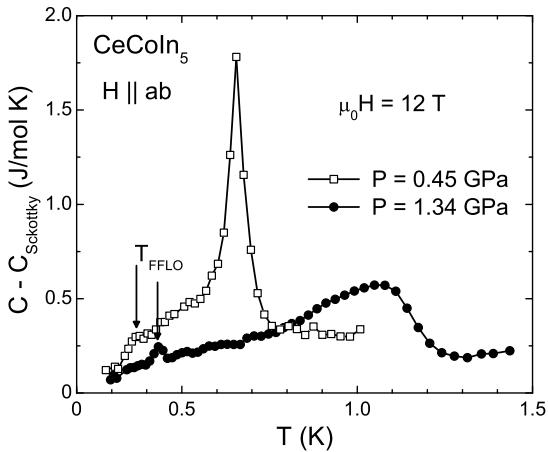}
\end{center}
\caption{Specific heat of CeCoIn$_5$ at $\mu_0H=12$ T ($H\parallel
ab$) for $P=0.45$ GPa and $P=1.34$ GPa. The nuclear Schottky
contribution to the heat capacity has been subtracted from the data.
The low temperature anomaly at $T_{\rm FFLO}$ is indicated by an
arrow.} \label{Cp_FFLO}
\end{figure}

\begin{figure}[tb]
\begin{center}
\includegraphics[angle=0,width=85mm,clip]{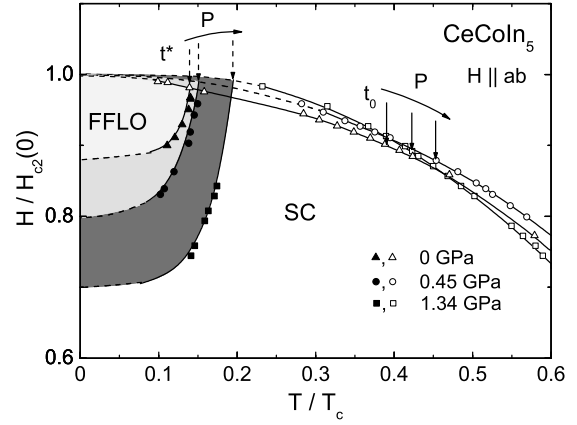}
\end{center}
\caption{Combined $H-T$ phase diagram of CeCoIn$_5$ for magnetic
field parallel to the $ab$-plane for 0, 0.45, and 1.34 GPa. The
magnetic field axis and the temperature axis are scaled by
$H_{c2}(0)$ and $T_c$, respectively. Open symbols indicate the
transition from the normal to the SC state, while solid symbols mark
the anomaly inside the SC state at $t_{\rm FFLO}=T_{\rm FFLO}/T_c$.
The tricitical point at $t^*=T^*/T_c$, where the FFLO transition
line hits the SC phase transition line, is marked by an dashed arrow
for each pressure, respectively. The crossover temperature from a
second-order to a first-order SC phase transition is indicated by a
solid arrow. $t^*$ as well as $t_0$ shift to higher temperature with
increasing pressure.} \label{PhD_scaled}
\end{figure}

Heat capacity measurements were carried out at 0, 0.45, and 1.34
GPa. The highest pressure was chosen in a way that the strong SF
present at ambient pressure are already substantially suppressed. In
the following we focus our discussion on the pressure evolution of
the $H-T$ phase diagram obtained for the magnetic field applied in
the $ab$-plane. For this orientation the existence of the anomaly
inside the SC state taken as a signature of the FFLO state has been
confirmed by different physical
probes.\cite{Radovan03,Bianchi03FFLO,Capan04,Watanabe04,Martin05}
Figure \ref{Cp_FFLO} shows specific heat data, $C-C_{\rm Schottky}$,
collected at $\mu_0H=12$~T for $P=0.45$~GPa and $P=1.34$~GPa. The
contribution of the nuclear Schottky anomaly to the specific heat at
low-$T$, $C_{\rm Schottky}$, has been subtracted from the data.
Below the SC transition a second anomaly appears (denoted as $t_{\rm
FFLO}=T_{\rm FFLO}/T_c$ in Fig. \ref{Cp_FFLO}) like the one observed
at atmospheric pressure. With increasing pressure, $T_{\rm FFLO}$ is
moving to higher temperatures. The evolution of the $t_{\rm
FFLO}(P)=T_{\rm FFLO}(P)/T_c$ phase line in the $H-T$ phase diagram
with pressure is depicted in Fig.~\ref{PhD_scaled}. In this phase
diagram $H$ is scaled by the respective $H_{c2}(0)$ and $T$ by $T_c$
for each pressure. With increasing pressure the FFLO phase is
expanding under pressure. It appears already at smaller reduced
fields, $h=H/H_{c2}(0)$: $h\approx0.88$ at ambient pressure and
$h\approx0.8$ and $h\approx0.7$ at 0.45~GPa and 1.34~GPa,
respectively. At a constant field, $t_{\rm FFLO}(h)$ shifts
continuously to higher temperature, similar to what is observed for
the tricritical point at $t^*$, where the extrapolated $t_{\rm
FFLO}(h)$ line meets the SC phase transition line. Note, that we did
not detect any additional anomaly for magnetic fields above the
upper-critical field at any pressure. Besides the second anomaly
inside the SC state we observe a remarkable change of both the shape
and the size of the anomaly at the SC transition upon increasing the
magnetic field. The mean-field-type shape at low fields sharpens up
and becomes more symmetrical at high magnetic fields, despite the SC
phase transition line being crossed at a glancing angle. This
indicates a crossover from a second-order to a first-order phase
transition, which we observe at all investigated pressures. The
crossover temperature $t_0=T_0/T_c$ is enhanced from $t_0=0.39$ at
atmospheric pressure to $t_0=0.42$ at 0.45~GPa and $t_0=0.45$ at
1.34~GPa, respectively. Our data are in good agreement with results
from magnetization studies under pressure.\cite{Tayama05} At
atmospheric pressure the $T_{\rm FFLO}$ anomaly is only observed
above the crossover field $H_0=H_{c2}(T_0)$ (see also Bianchi {\it
et al.}\cite{Bianchi03FFLO}). This is not the case under applied
pressure anymore. Here, the SC anomaly still displays a
mean-field-type shape, indicating a second-order phase transition,
while the $T_{\rm FFLO}$ anomaly is already appearing at lower $T$
inside the SC state. The SC transition is of first order only at
fields greater than $H_0$, much higher than the field where the FFLO
anomaly first occurs. Figure \ref{Cp_FFLO} shows data at
$P=0.45$~GPa and $\mu_0H=12$~T where the transition to the SC state
is of first order ($H_0=11.2$~T/$\mu_0$), while for $P=1.34$~GPa at
the same field the transition still displays a mean-field-type shape
($H_0=12.4$~T/$\mu_0$).

\section{Discussion}

In the following we want to verify, if the necessary conditions for
the formation of the FFLO state are still fulfilled in CeCoIn$_5$
under pressure: i) clean-limit type II superconductor and ii)
Pauli-limited with a Maki-Parameter $\alpha>1.8$. Finally, we will
discuss the influence of AFM-SF on the FFLO phase.

\begin{figure}[tb]
\begin{center}
\includegraphics[angle=0,width=75mm,clip]{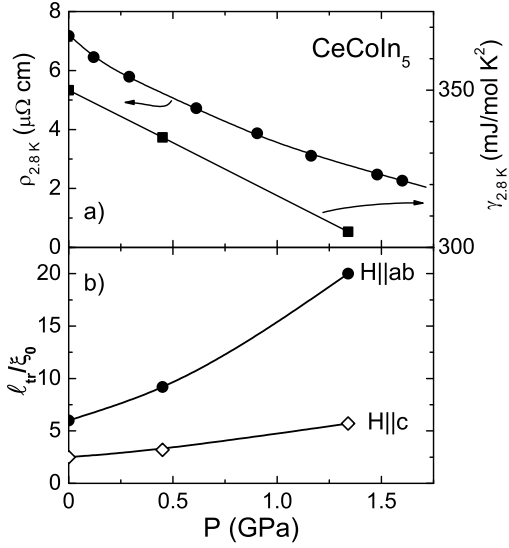}
\end{center}
\caption{Upper panel: left axis, resistivity at 2.8 K,
$\rho_{\rm\,2.8 K}$ (taken from Nicklas {\it et al.}
\cite{Nicklas01}) and, right axis, Sommerfeld-coefficient at 2.8 K,
$\gamma_{\rm 2.8\,K}=C/T({\rm 2.8\,K})$ as function of pressure;
lower panel: ratio of estimated mean-free-path, $\ell_{tr}$, and
coherence length, $\xi_0$, for $H\parallel ab$ and $H\parallel c$ as
function of pressure. See text for details.} \label{rho_gamma}
\end{figure}

The ratio of the quasiparticle mean-free path and the coherence
length, $\ell_{tr}/\xi_0$, increases significantly with pressure for
the magnetic field applied in the basal plane as well as for the
field perpendicular to it (see Fig. \ref{rho_gamma}, lower panel).
CeCoIn$_5$ becomes even {\it cleaner} with pressure. An upper limit
for $\xi_0$ can be obtained using the BCS relation
$\xi_0=\sqrt{\Phi_0/(2\pi \mu_0H_{c2}(0))}$, where $\Phi_0$ is the
flux quantum. For the estimation of $\ell_{tr}$ we follow the scheme
of Orlando {\it et al.}\cite{Orlando79}, where the quasiparticle
mean-free path is given by $\ell_{tr}\propto
1/(\xi_0T_c\gamma_{n}\rho_{n})$. Here, $\gamma_{n}$ and $\rho_{n}$
are the Sommerfeld-coefficient and the resistivity in the normal
state right above the SC transition, respectively. $\ell_{tr}$
increases substantially with increasing pressure. $\gamma_{\rm
2.8\,K}=C({\rm 2.8\,K})/T$ decreases only slightly, from about 350
mJ/mol K$^2$ at atmospheric pressure to 305 mJ/mol K$^2$ at
1.34~GPa. The main effect on $\ell_{tr}$  originates from the strong
reduction of the resistivity in the normal state (see Fig.
\ref{rho_gamma}, upper panel), which is caused by a pressure-induced
change of the inelastic scattering rate reflecting the reduction of
the AFM-SF as expected for an increasing distance to the
QCP\cite{Sidorov02}.

\begin{figure}[tb]
\begin{center}
\includegraphics[angle=0,width=85mm,clip]{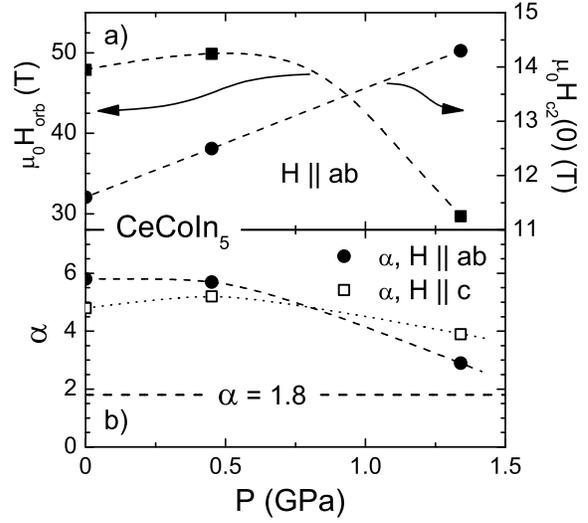}
\end{center}
\caption{Upper panel: left axis, orbital-limiting field, $H_{\rm
orb}$ and, right axis, upper-critical field, $H_{c2}(0)$ versus
pressure; lower panel: Maki-parameter, $\alpha=\sqrt{2}H_{\rm
orb}/H_P$, for $H\parallel ab$ and $H\parallel c$ as function of
pressure.} \label{HorbHc2Alpha}
\end{figure}

Pauli paramagnetism leads to an upper limit for the magnetic field
which can be still supported by the superconductor, the Clogston
paramagnetic limit $H_{\rm P}=\Delta_0/\sqrt{2}\mu_{B}$
\cite{Clogston62} with $\Delta_0$ the SC energy gap and $\mu_{B}$
the Bohr magneton. In addition, orbital effects also limit $H_{c2}$.
Maki and Tsuneto\cite{Maki64} showed that for a Pauli-limited system
($\alpha\geq1$) the second order transition from the paramagnetic to
the mixed state becomes instable, and for
$\alpha\longrightarrow\infty$ a first order transition is expected
below $t_0=T_0/T_c=0.56$\cite{Maki64}. The crossover from a
second-order to a first-order phase transition observed at all
pressures in CeCoIn$_5$ indicates that the strong Pauli-limiting
effect is present also under pressure. The experimentally obtained
$t_0$ is always smaller than the theoretical threshold
$t_0=T_0/T_c=0.56$.

The Maki parameter $\alpha=\sqrt{2}H_{orb}/H_P$ can be estimated
directly. The orbital-limiting field $H_{\rm orb}=0.7\,T_c(\partial
H_{c2}(T)/\partial T)|_{T=T_c}$\cite{Helfand,Hake67} is easily
accessible, while for the Pauli limiting field $H_{\rm
P}=H_c/(2\sqrt{\pi\chi_{\rm spin}})$ the knowledge of the the spin
susceptibility $\chi_{\rm spin}$ is needed. $H_c$ is the
thermodynamic critical field. Without knowing $\chi_{\rm spin}$, we
use a reasonable approximation $H_{\rm P}=H_{\rm c2}$. A more
detailed analysis of the $H_{c2}(T)$ data within BCS theory gives
values for $H_{\rm P}$ close to the approximation used
here.\cite{Miclea06,Won04} The orbital-limiting field is strongly
reduced on increasing the pressure, but is still about twice as
large as $H_{c2}(0)$ at 1.34 GPa. This indicates that Pauli limiting
still is the dominating effect in limiting the upper-critical field
(see Fig.~\ref{HorbHc2Alpha}). The Maki parameter decreases for
$H\parallel ab$ from $\alpha=5.8$ at atmospheric pressure to
$\alpha=2.9$ at 1.34~GPa, which is still larger than the minimum
value of $\alpha=1.8$ required for the realization of the FFLO state
in an $s$-wave superconductor.\cite{Gruenberg66}

On applying pressure Pauli-limiting becomes weaker, but still
exceeds the orbital-limiting effects. Furthermore, the crossover
from a first-order to a second-order SC to normal phase transition
at $H_0=H_{c2}(T_0)$ is expected only in the case of a strongly
Pauli limited upper critical field. From our observation that under
pressure i) the Maki parameter $\alpha>1.8$, ii) the ratio
$\ell_{tr}/\xi_0$ is increased, and iii) the anomaly at $T_{\rm
FFLO}$ inside the SC state still exists, we conclude that the FFLO
state is indeed formed in CeCoIn$_5$. However, the microscopic
realization of the FFLO state in CeCoIn$_5$ needs further
exploration. We found the FFLO phase expanding, although the Maki
parameter decreases and while the spin fluctuations are strongly
suppressed with increasing pressure. This suggests that SF have a
strong detrimental effect on the FFLO phase, in agreement with
theoretical predictions.\cite{Adachi03} Recent NMR studies at
atmospheric pressure suggest the existence of local moment magnetism
inside the normal vortex cores,\cite{Young06} in contrast to earlier
reports.\cite{Kakuyanagi05} This study highlights the intricate
relation of magnetism and the FFLO phase close to $H_{c2}(0)$,
consistent with the NFL behavior observed in the normal state close
to $H_{c2}(0)$. Our data reveal that the expansion of the FFLO phase
with increasing pressure is consistent with the strong suppression
of the AFM-SF fluctuations.\cite{Sidorov02,Yashima04} In fact LFL
behavior is recovered at high
pressure.\cite{Sidorov02,Lengyel02,Shishido03,Yashima04} In
experiments carried out under pressure we would expect a drastic
weakening of the spin polarization observed in the NMR measurements
inside the FFLO phase at atmospheric pressure.

\section{Conclusions}

We have found that an anomaly at $T_{\rm FFLO}$ inside the SC phase
in CeCoIn$_5$ close to $H_{c2}(0)$  established
previously\cite{Radovan03,Bianchi03FFLO,Capan04,Watanabe04,Martin05}
persists upon applying hydrostatic pressure. Also, the conditions
for the formation of the FFLO state, namely Pauli-limiting exceeding
orbital limiting and clean-limit SC, are found to be still fulfilled
under pressure. While the Pauli-limiting effect is becoming weaker
compared with orbital limiting on applying pressure, the FFLO phase
is expanding, in contrast to what is naively expected. Considering
the strong suppression of the AFM-SF on increasing pressure, the
expansion of the FFLO phase can be explained by a strong detrimental
effect of the AFM-SF on the FFLO state.

\section*{Acknowledgments}
We thank R. Borth and R. Koban for machining the pressure cell and
for their help in the preparation process. We also thank A. D.
Bianchi, E. Lengyel, R. Ikeda, K. Maki, A. C. Mota, D. Parker, and
P. Thalmeier for stimulating discussions. This work was in part
supported by the DFG under the auspices of the SFB 463. Work at Los
Alamos National Laboratory was performed under the auspices of the
U.S. Department of Energy.

\end{document}